\documentclass[cits]{PoS}
\usepackage{url}
\title{The TANAMI Program: Southern-Hemisphere AGN on (Sub-)parsec Scales}

\ShortTitle{The TANAMI Program: Southern-Hemisphere AGN on (sub-)parsec Scales}

\author{\speaker{Cornelia M\"uller}, F.~Krau\ss{}\\%
        Dr. Karl Remeis-Observatory \& ECAP, 96049 Bamberg, Germany\\
	 Julius-Maximilian-Universit\"at W\"urzburg, 97074 W\"urzburg, Germany\\
        E-mail: \email{cornelia.mueller@sternwarte.uni-erlangen.de}}

\author{M.~Kadler, J.~Tr\"ustedt\\
       Julius-Maximilian-Universit\"at W\"urzburg, 97074 W\"urzburg, Germany}

\author{R.~Ojha\\
       NASA Goddard Space Flight Center, Greenbelt, MD 20771, USA}

\author{E.~Ros\\
Dept. d'Astronomia i Astrof\'{\i}sica, Universidad de Val\`encia, 46100 Burjasot, Spain\\
 Max-Planck-Institut f\"ur Radioastronomie, Bonn, Germany}  

\author{J.~Wilms\\                                                                                                                                                                
          Dr. Karl Remeis-Observatory \& ECAP, 96049 Bamberg, Germany}  

\author{M.~B\"ock\\
        Max-Planck-Institut f\"ur Radioastronomie, 53010 Bonn, Germany}

\author{M.~Dutka, B.~Carpenter\\
       The Catholic University of America, Washington, DC 20064, USA}
   
\author{and the TANAMI collaboration}

\abstract{
  The Southern Hemisphere VLBI monitoring program TANAMI provides dual-frequency (8\,GHz and 22\,GHz), 
  milliarcsecond monitoring of extragalactic jets south of $-30^\circ$ declination.
  The TANAMI sample consists of a 
  combined radio and $\gamma$-ray selected subsample of currently $\sim 80$ AGN jets, with new $\gamma$-ray bright sources being added 
  upon detections 
  by \textsl{Fermi}/LAT. 
  Supporting programs provide 
  simultaneous multiwavelength coverage of all sources, in order to construct broadband spectral energy distributions 
  (SEDs) of flaring and quiescence source states, as well as a rapid follow-up of high-energy flares. 
  This combined setup allows us to continuously study the spectral and structural evolution of highly energetic extragalactic jets
  and test correlations in different wavebands,
  providing crucial information on underlying physical mechanisms.
  Here, we present jet kinematics of Centaurus~A and show preliminary VLBI results on 
  PKS\,0625--354 and the time-dependent spectral index image of PKS\,0537--441.
   }

\FullConference{11th European VLBI Network Symposium \& Users Meeting,\\
		October 9-12, 2012\\
		Bordeaux, France}

\begin{document}
\section{Introduction}
The basic concept of the TANAMI\footnote{\url{http://pulsar.sternwarte.uni-erlangen.de/tanami/}} program \cite{Ojha2010a} consists of a frequent (with every source observed approximately every three months)
Very Long Baseline Interferometry (VLBI) monitoring of a combined radio- and $\gamma$-ray selected sample of $\sim 80$ southern extragalactic jets.  Simultaneous 8\,GHz and 22\,GHz observations are performed using the Australian Long Baseline Array (LBA) and associated telescopes in Antarctica, 
Chile, New Zealand and South Africa. 
TANAMI is the only large VLBI monitoring program with such a simultaneous dual-frequency approach,
allowing us to study the evolution of the jet morphology and, in addition, the variation of the spectral index 
along the jet at highest spatial resolution.
For about a third of the source sample, TANAMI provides the first ever VLBI images \cite{Mueller2012a}.
This is completed with contemporaneous multiwavelength observations from an intense radio
monitoring campaign with ATCA and single dish observations with Ceduna, X-ray coverage with \textsl{Swift} and \textsl{XMM-Newton},
and continuously observing by \textsl{Fermi}/LAT in $\gamma$-rays (see \cite{Krauss2013} for more information).

With this setup, TANAMI aims to study the broadband emission behavior of jets at different source states. Our dual-frequency VLBI observations provide 
insight into structural and spectral changes of the jets at milliarcsecond (mas) resolution.
In order to investigate the jet emission mechanism producing highly energetic photons and the putative
link to the structural changes seen at parsec scales (also known as the `radio-gamma-connection in jets'),
we perform a combined multiwavelength study of the spectral energy distributions (SEDs) and the VLBI morphology of TANAMI sources
since the end of 2007. 
We report on recent results of the VLBI analysis, showing first jet kinematics and time evolution 
of the spectral index. The multiwavelength analysis is described in detail in \cite{Krauss2013}.

\section{Jet Kinematics}
Since the end of 2007, TANAMI is frequently performing dual-frequency observations of all sources
in order to study changes in the jet structure, like jet component ejections.
The $(u,v)$-coverage of the TANAMI array is considerably better than that of former 
Southern Hemisphere VLBI measurements with long baselines to the IVS antennas
GARS/O'Higgins (Antarctica) and TIGO (Chile) and to the contribution of the new Warkworth
antenna (New Zealand) since 2011. 
As of 2012, for almost the whole TANAMI sample the
minimum required number of 8\,GHz observations became
available to perform first kinematics analysis.
In order to build up a consistent and reliable kinematic model of the jet flow, experiences of other VLBI
monitoring programs (e.g., MOJAVE \cite{Lister2009a}) show, that at least five observation epochs should be considered. Therefore, the TANAMI
monitoring allows us to start now with the jet kinematic analysis of the initial source 
sample \cite{Ojha2010a} for which sufficient time coverage is available. 

Here, we present first results of the ongoing analysis on the jet kinematics of the jets of Centaurus~A and PKS\,0625--354 at milliarcsecond scales.
\subsection{Centaurus~A's Jet Kinematics at Sub-parsec Scales}
At a distance of only 3.8\,Mpc \cite{Harris2010}, Centaurus~A is the closest radio-loud AGN. 
The TANAMI array provides an angular resolution of less than 1\,mas (which translates into 0.018\,pc),
resulting in the highest resolved image of an extragalactic jet-counterjet system resolving the jet down to 
sub-parsec scales \cite{Mueller2011a}. 
The detailed spectral index map shows multiple possible regions of $\gamma$-ray emission detected by \textsl{Fermi}/LAT \cite{Mueller2011a, Abdo2010d} challenging single-zone broadband emission models.

\begin{figure}[!ht]
    \centering
\includegraphics[height=9cm]{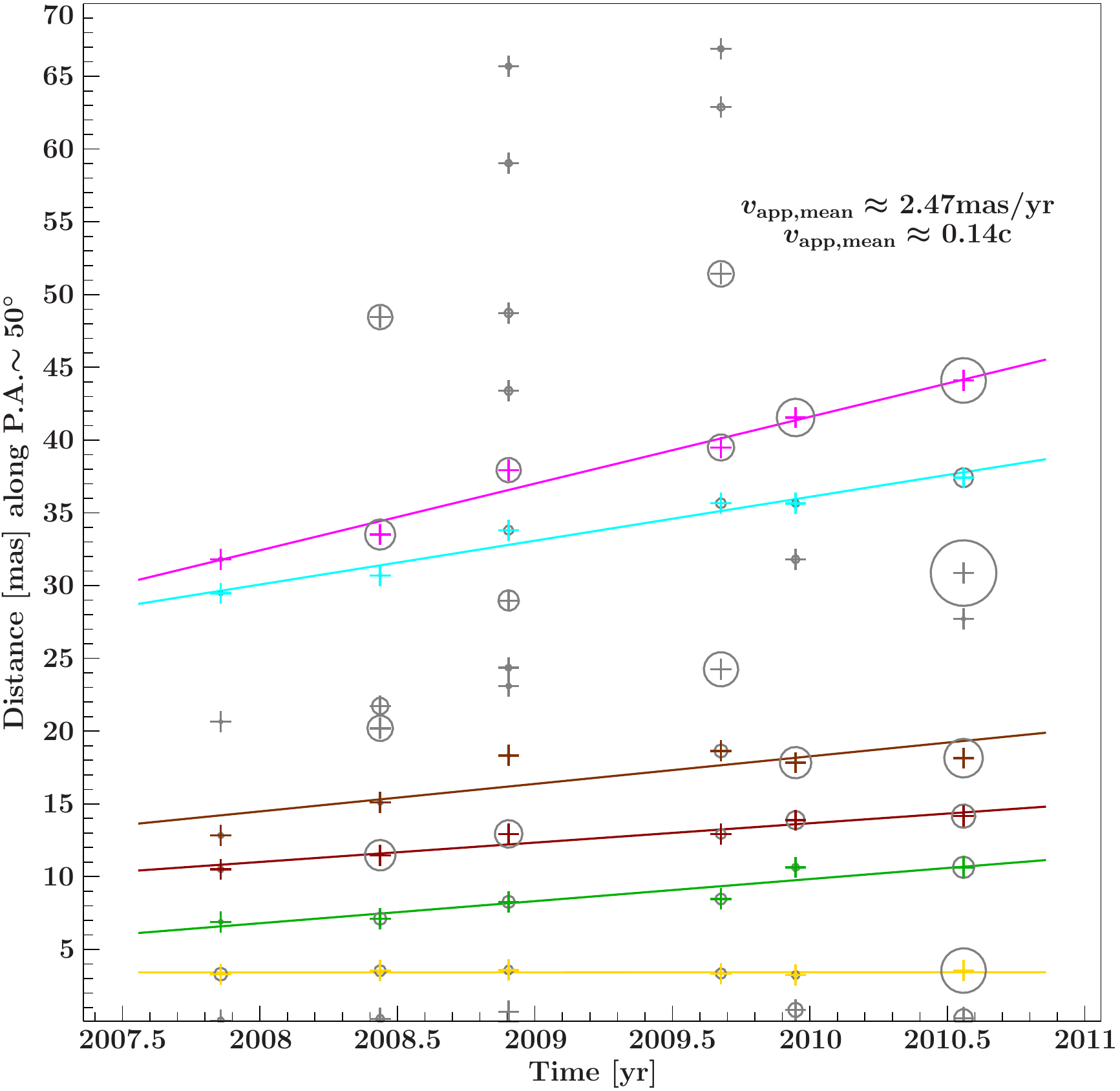}\hfill
\caption{Jet kinematic analysis of Centaurus~A. 
  Separation of individual, associated jet components along the jet axis at a jet position angle of $50^\circ$ 
  with time for the first six TANAMI observations.
  Crosses mark the central position of every component with the FWHM of the Gaussian model components
  shown as gray circles.
  Straight lines indicate the corresponding linear regression fits to the distance 
  giving in a mean apparent jet speed of $v_\mathrm{app}\propto 2.47$mas/yr\,$\propto 1.4\,c$ (yellow indicates the stationary component at $\sim$3.5\,mas).
  }
\label{fig1}
\end{figure}

The analysis of the jet evolution over time (see Fig.~\ref{fig1}, M\"uller et al., in prep.) gives a very complex jet morphology with 
varying jet speeds for individual identified components. 
By fitting Gaussian model components to the $(u,v)$-data, we can identify and track 
up to eight bright, distinct jet features resulting in apparent speeds ranging 
from $v_\mathrm{app}\propto 1$mas/yr to $v_\mathrm{app}\propto 4$mas/yr.

A stationary component, potentially a standing shock, is found downstream at a distance of $\sim$3.5\,mas next to the jet core. This bright
component
has a flat spectrum \cite{Mueller2011a} and possibly corresponds to the quasi-stationary core 
extension seen in previous VLBI observations \cite{Tingay2001b}.
Further downstream ($\sim$25\,mas) a jet widening is present, showing a substantial drop
in surface brightness and subsequent collimation is detected at 8\,GHz, co-spatial with an
emission feature at 22\,GHz \cite{Mueller2011a}. 
Over three years of TANAMI observations, this peculiar feature appears to be stationary, while
the speed of individual components ahead and after it shows no significant change. 

The mean apparent speed of all moving components of $v_\mathrm{app}\propto 2.47$mas/yr\,$\propto 1.4\,c$
gives an estimate for an underlying, continuous jet flow and is consistent with
previous results \cite{Tingay2001b}.

\subsection{Superluminal Motion in PKS\,0625--354}
\begin{figure}[!ht]
\centering
\includegraphics[height=6.5cm]{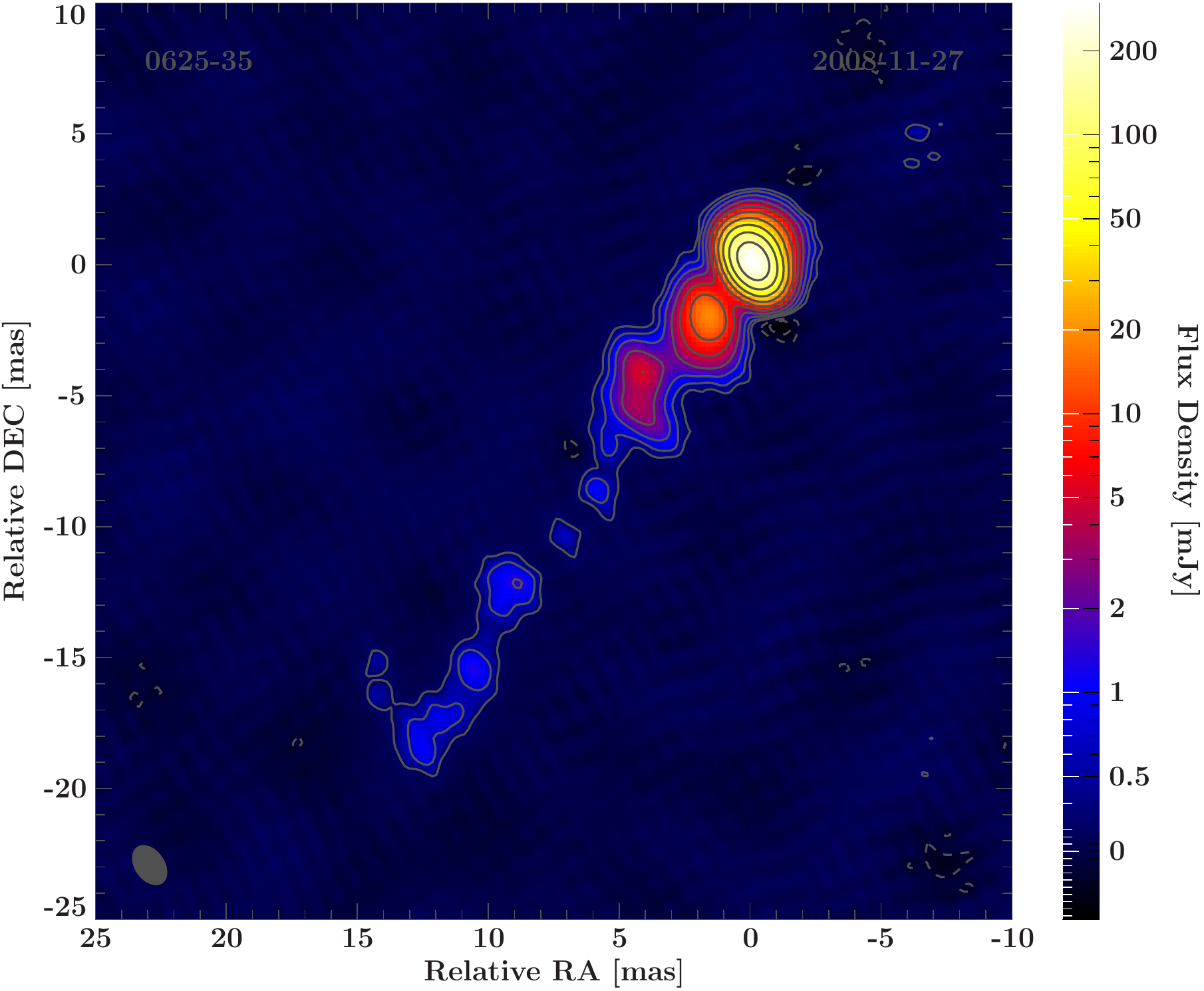}\hfill
\includegraphics[height=6.5cm]{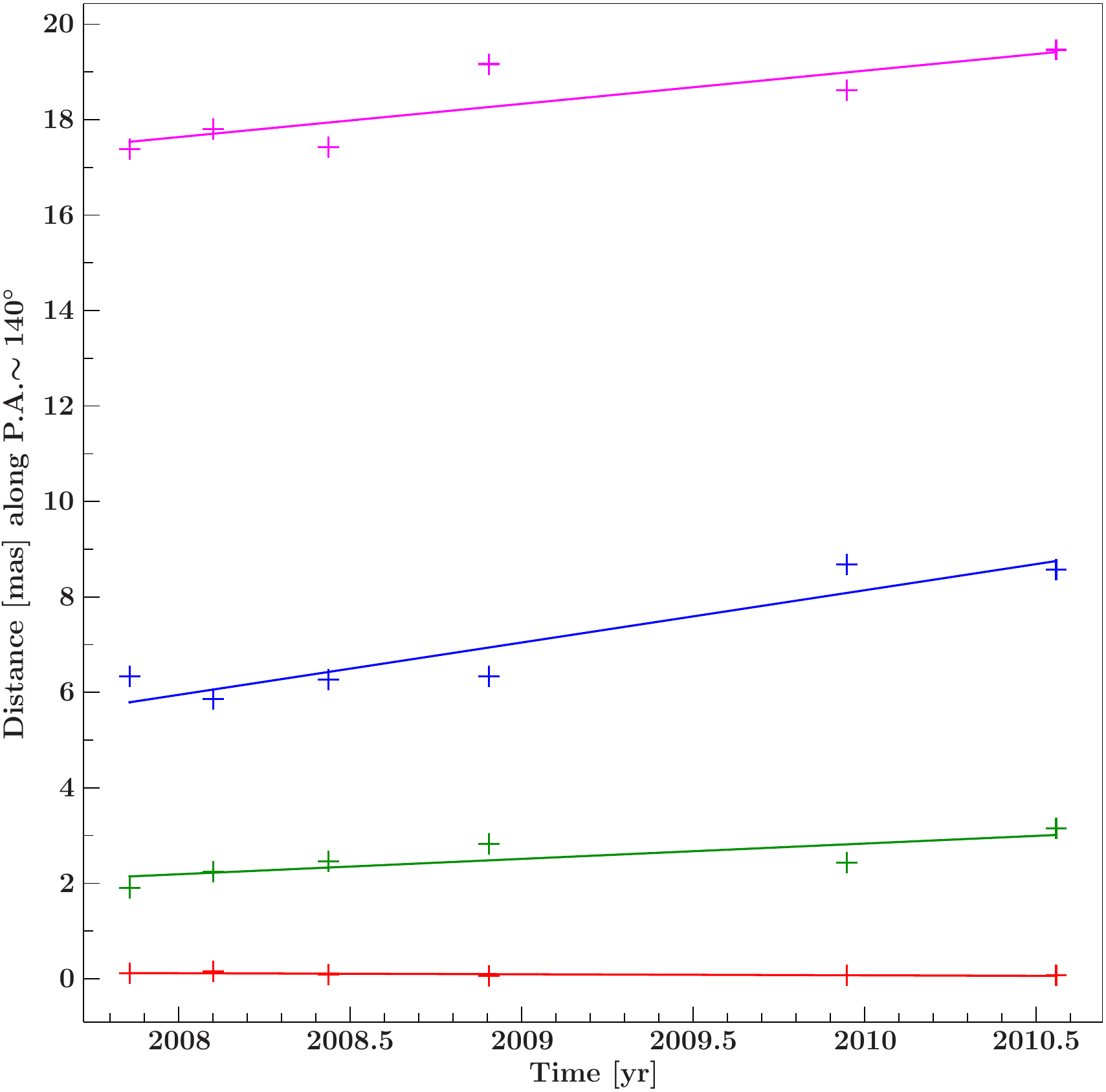}
\caption{
\textit{Left:} TANAMI image of PKS\,0625--354 at 8\,GHz (2008 November). Contours are logarithmic, 
separated by a factor 2, with the lowest level set to the 3$\sigma$-noise-level.
\textit{Right:} Preliminary kinematic analysis of the inner pc-scale jet of  PKS\,0625--354 revealing a mean apparent speed 
of $v_\mathrm{app}\propto 0.7$\,mas/yr\,$\propto 3c$. The plot shows the jet component separation over time 
of three identified jet components of the first six TANAMI observations at 8\,GHz. 
The innermost component (red) was identified as the core and was taken as the reference point for component position.
Linear regression fits were performed to derive component speeds.
}
\label{fig3}
\end{figure}

The misaligned radio galaxy PKS\,0625--354 ($z=0.55$) has a large scale FR~I structure \cite{Morganti1999}, but it is also classified as a LINER \cite{Lewis2003}. 
It has been detected by \textsl{Fermi}/LAT \cite{Abdo2010c, Nolan2012} as a high luminosity and hard spectrum source, supporting a  BL Lac classification
\cite{Wills2004}.
Our TANAMI observations revealed a single-sided, parsec-scale jet morphology 
(Fig.~\ref{fig3}). We identify the jet core as the optically thick 
and brightest component and use it as the reference point to determine the separation of each component from the core over time. 
The preliminary kinematic analysis results in a superluminal motion of $v_\mathrm{app}\propto3.0\pm 0.5c$ (Tr\"ustedt et al., in prep.) .

\section{Parsec-scale Spectral Index Imaging}

\begin{figure}
\centering
\includegraphics[width=0.33\textwidth]{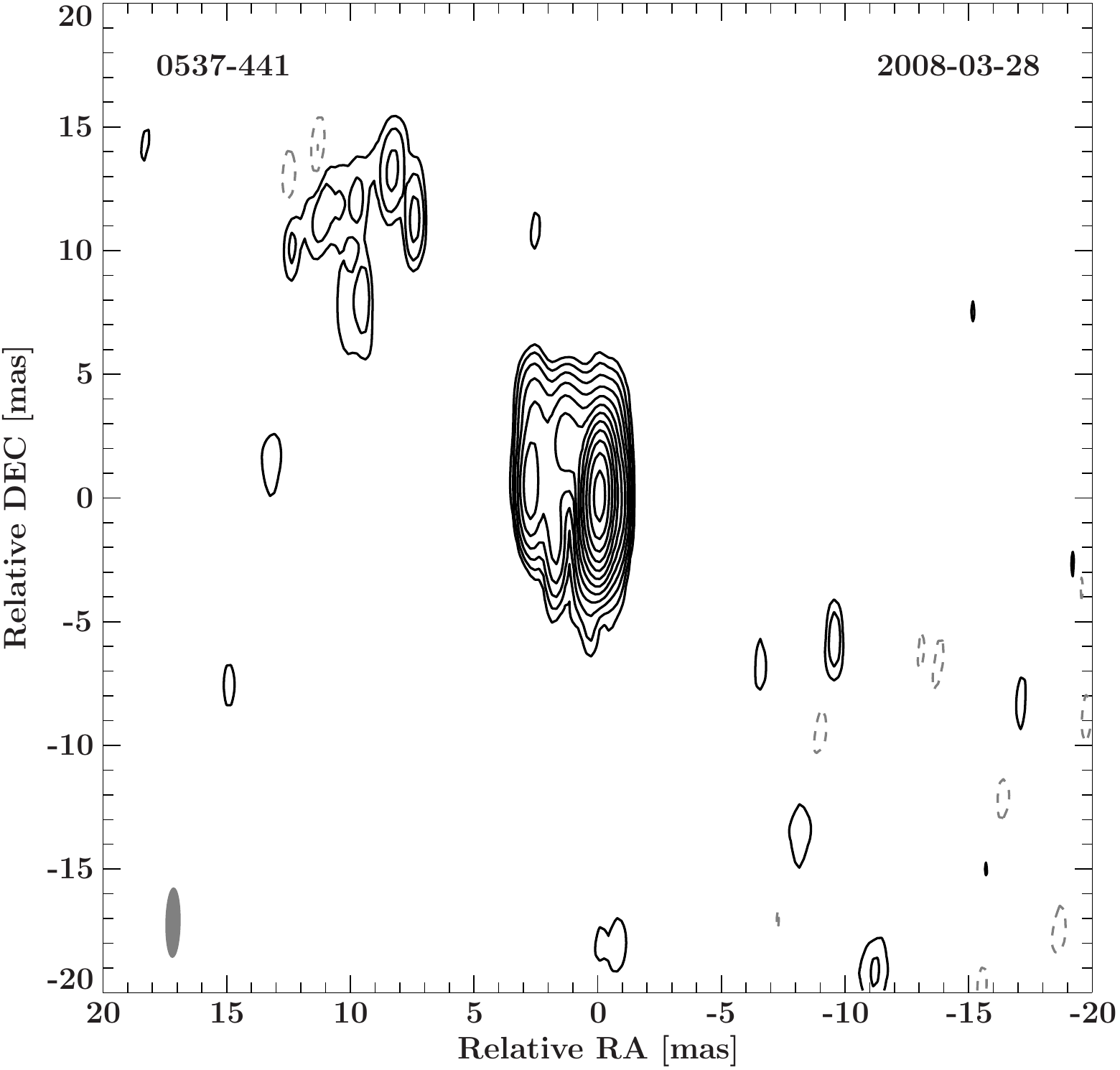}\hfill
\includegraphics[width=0.33\textwidth]{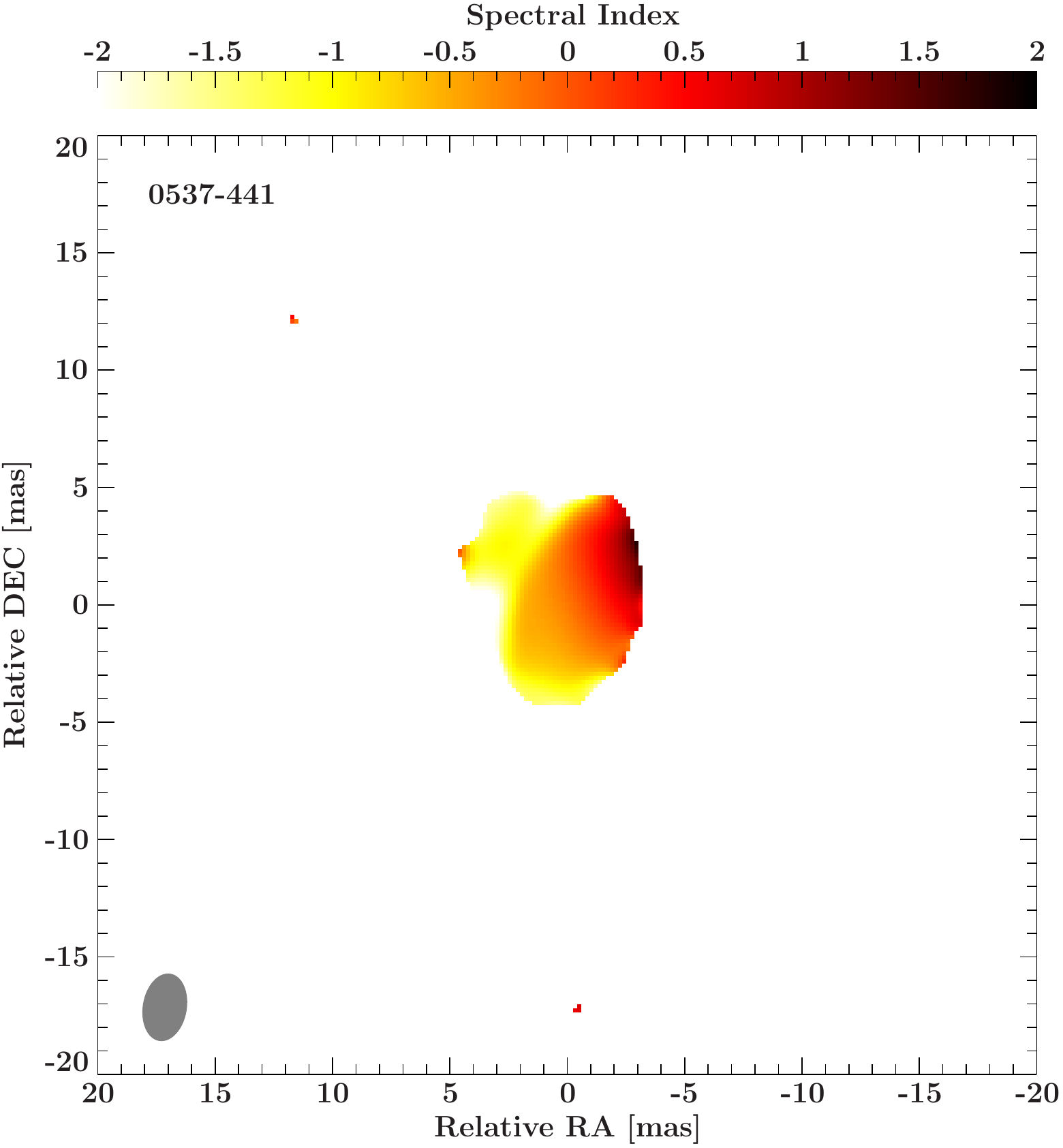}\hfill
\includegraphics[width=0.33\textwidth]{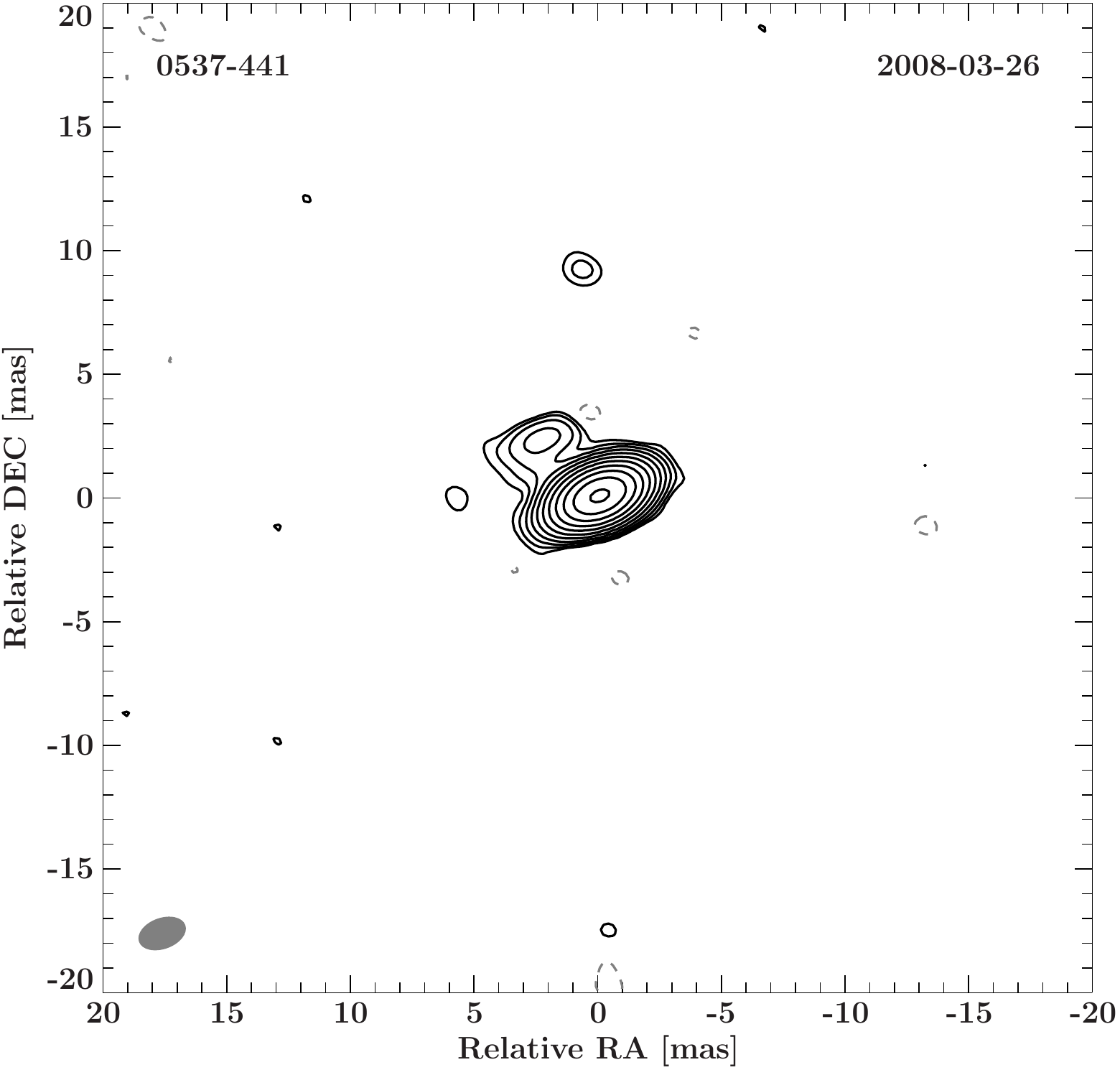}
\includegraphics[width=0.33\textwidth]{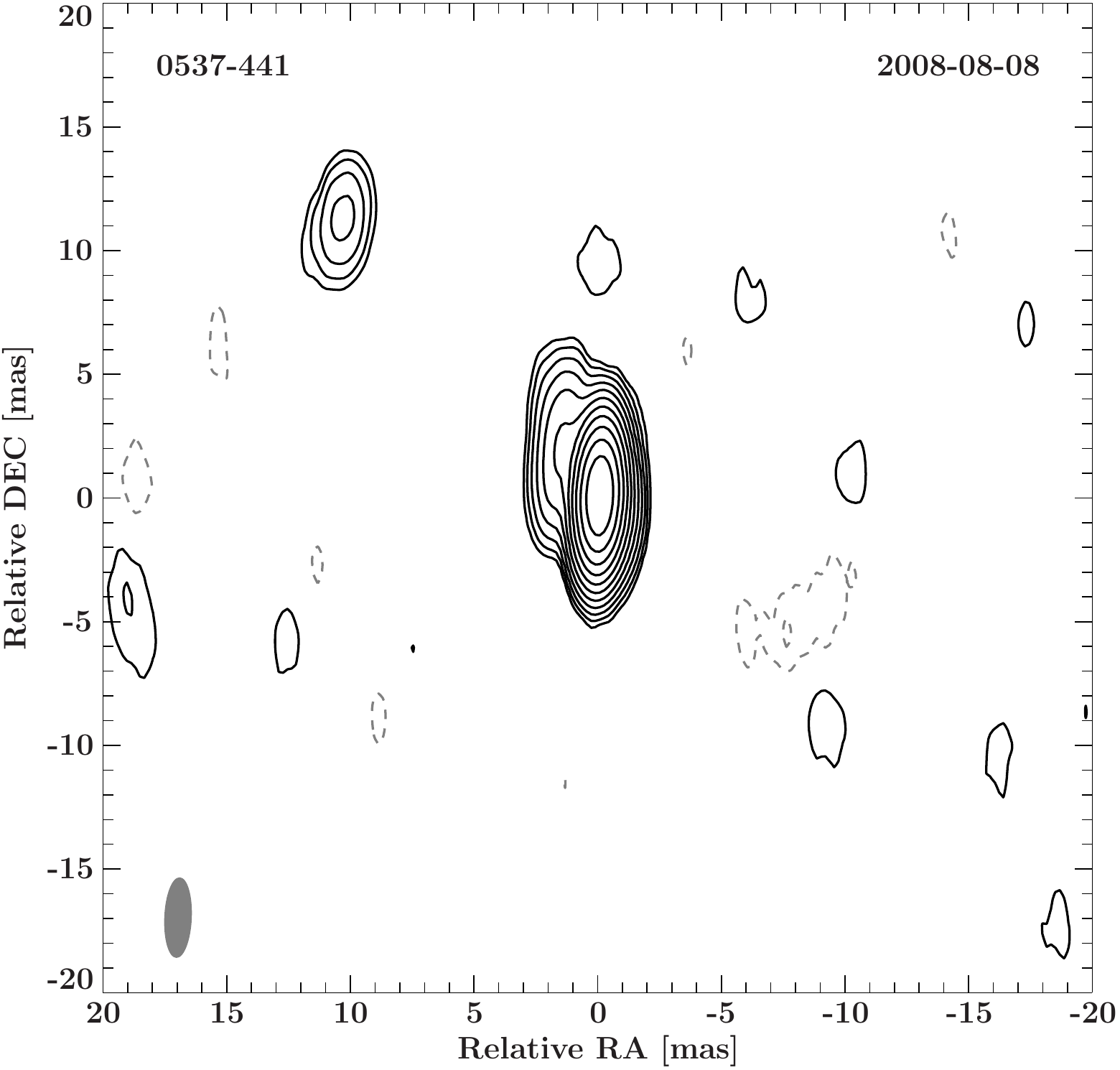}\hfill
\includegraphics[width=0.33\textwidth]{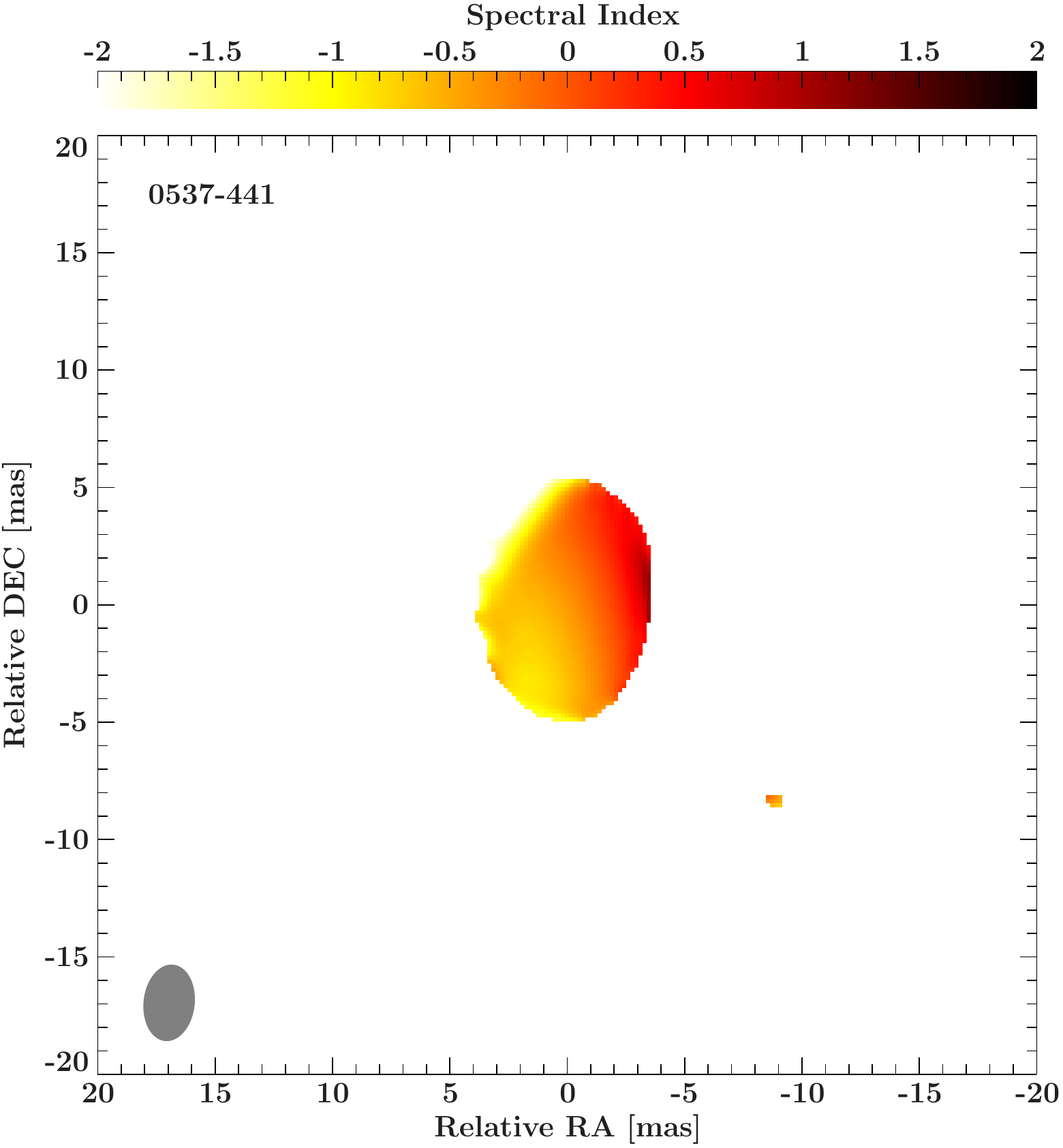}\hfill
\includegraphics[width=0.33\textwidth]{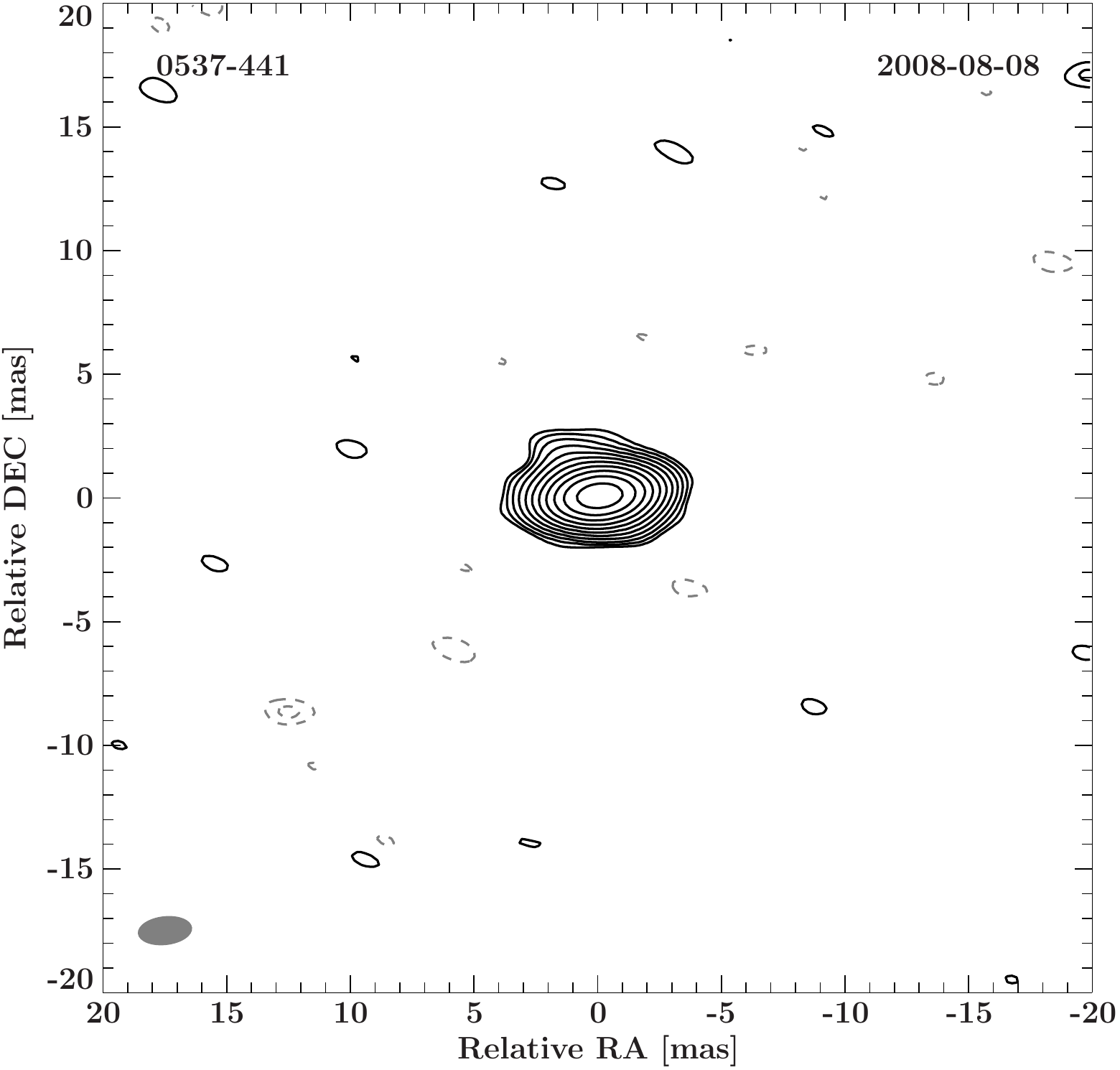}
\caption{Two simultaneous dual-frequency observations of PKS\,0537--441 in March 2008 (top) and August 2008 (bottom).
  The left and right panels show the 8\,GHz and 22\,GHz images, respectively, with the corresponding spectral index map in the middle panel.
  The color scale indicates the spectral index distribution: optically thick regions appear darker.
  Contours in the left and right panels are logarithmic, separated by a factor 2, with the lowest level set to the 3$\sigma$-noise-level.
  The spectral index is derived where the flux densities at 8\,GHz and 22\,GHz exceed both the 3$\sigma$-noise-level. 
  To construct the spectral index map both images have been restored with a common beam represented by the gray ellipse in the lower left corner.}
\label{fig4}
\end{figure}

The quasi-simultaneous dual-frequency VLBI monitoring at 8\,GHz and 22\,GHz enables us to construct images of the spectral distribution  
(with spectral index $\alpha$ defined as $F_\nu \propto \nu^{+\alpha}$) along the jet with high angular resolution. 
This can be used to identify possible $\gamma$-ray production sites (e.g., \cite{Mueller2011a}) as 
the high energy emission is expected to originate from the optically thick regions within the
jet.
The time-dependent study of the spectral index maps provides insight into the spectral changes
of individual features along the jets resolving possible active regions. 
Combined with multiwavelength investigations at different source states (flaring,
quiescence) this spectral monitoring at very high angular resolution is a powerful tool
to test different jet emission models \cite{Krauss2013}.
As a representative example for such  spectral index maps, Fig.~\ref{fig4} shows the 8\,GHz, 22\,GHz and corresponding
spectral index map for two simultaneous TANAMI observations of the TeV-blazar 
PKS\,0537--441.

As described in more detail in \cite{Krauss2013}, the broadband SED of PKS\,0537--441 shows significant 
differences for different source states. The highly resolved dual-frequency images and spectral index 
maps provided by TANAMI observations will play an important role in distinguishing between possible broadband emission models. 

\section{Conclusion and Outlook}
TANAMI is an ongoing VLBI monitoring program studying the kinematic and spectral changes
of Southern extragalactic jets with mas resolution. Jet component ejection events and (radio) spectral changes
are being determined in order to investigate possible correlations with high-energy 
flares and to study the broadband SEDs of the $\gamma$-ray
bright subsample at different source states.

We started intense combined VLBI and multiwavelength studies on individual
sources, testing different theoretical
emission models for blazar SEDs. 
These try to
explain the broadband emission seen from radio-loud
AGN with leptonic or hadronic (or a combination of
both) processes. 
As an important factor, the high resolution TANAMI observations
allow us to distinguish between core and total jet emission.
Quasi-simultaneous multiwavelength
monitoring of flaring and quiescence states of the whole 
source sample will help us to observationally determine
which model is most likely to be correct. 

\acknowledgments{This work has been partially supported by the Studienstiftung
des deutschen Volkes through a fellowship
to C. M\"uller and by the European Commission under
contract ITN 215212-'Black Hole Universe'. E. Ros
acknowledges partial support by the Spanish government
through grant AYA2009-13036-C02-02 and by
the COST action MP0905 `Black holes in a violent
Universe'.
}

\end{document}